\documentclass[11pt,oneside,english]{article}
\usepackage[a4paper, total={6.2in, 9in}]{geometry}
\usepackage[utf8]{inputenc}
\pdfoutput=1

\usepackage{rotating}
\usepackage{graphicx}
\usepackage{amssymb}
\usepackage[table]{xcolor}
\usepackage{xcolor}
\usepackage[T1]{fontenc}
\usepackage{lmodern}

\usepackage{multirow,booktabs,array,longtable,caption}

\newcommand{\changes}[1]{#1}

\newcommand{\nochange}[1]{#1}

\usepackage{enumerate}

\usepackage{caption}

\usepackage{courier}  

\usepackage[bottom, multiple]{footmisc}  
\usepackage[normalem]{ulem}    
\usepackage{anyfontsize}

\PassOptionsToPackage{hyphens}{url}\usepackage[hidelinks, hyperfootnotes=false]{hyperref}
\usepackage{fancyhdr}                           
\fancypagestyle{firstpage} 
{
   \footskip=8mm

   \fancyhf{}

   \fancyfoot[C]{\vspace{-6mm}\thepage}

   \fancyfoot[L]{\vspace{-1mm}\noindent\rule{6.2in}{0.4pt}
     \scriptsize{
       \changes{*This paper appeared in the Journal of Financial Market Infrastructures,
       Vol 10, No 4 (June 2022), pages 55-63.} \newline
         \textcopyright \hspace{0.5mm} \changes{Barclays 2022-2023} \\
       This work is licensed under a \href{https://creativecommons.org/licenses/by/4.0/}{Creative Commons Attribution 4.0 International License (CC
       BY).}  \\ 
       Provided you adhere to the CC BY license, including as to attribution, you are
       free to copy and redistribute this work in any medium or format and remix, transform,
       and build upon the work for any purpose, even commercially.                  \\
       \changes{BARCLAYS is a registered trademark,} all rights are reserved. \\
     }
 }
}

\title{\vspace{-2cm}\changes{Illustrative} Industry Architecture \\
to Mitigate Potential Fragmentation across \changes{a}\\
Central Bank Digital Currency and Commercial Bank Money*}

\author{%
  \begin{tabular}{c} {\fontsize{10.75}{1cm}\selectfont Lee Braine and Shreepad Shukla} \\ {\fontsize{10.75}{1cm}\selectfont Chief Technology Office} \\
    {\fontsize{10.75}{1cm}\selectfont Barclays} \\ \hskip 1em \end{tabular} }

\date{\vspace{-0.5cm}March 31, 2022}

\begin{document}
\maketitle

\vspace{-1.40cm}
\begin{center}{\fontsize{8}{1cm}\selectfont \changes{(Revised September 20, 2023)}}\end{center}

\thispagestyle{firstpage} 
\vspace{-0.75cm}
\begin{abstract}
\noindent
Central banks are actively exploring central bank
digital currencies (CBDCs) by conducting research, proofs of concept and
pilots.
However, adoption of a CBDC can risk fragmenting both payments markets and
retail deposits.
In this paper, we aim to provide a mitigation to this fragmentation risk
by presenting an illustrative industry architecture \changes{that}
places CBDCs and commercial bank money on a similar footing.
We introduce the concept of ecosystems providing a common
programmability layer that interfaces with the account systems at both
commercial banks and the central bank.
We focus on a potential \changes{UK} CBDC, including
industry ecosystems interfacing with commercial banks
using \nochange{Open Banking} application programming interfaces.

\end{abstract}

\vspace{0.25cm}

\section{Introduction}
\label{sec:introduction}

\noindent
A central bank digital currency (CBDC) is a digital payment instrument, denominated
in a national unit of account, that is a direct liability of a central 
bank \cite{bis-aer2021-cbdcs}.
Central banks are actively exploring CBDCs \cite{bis-cbdc-survey3, ac-cbdc-tracker} with
various motivations such as:

\begin{enumerate}[(i)]
  \item \changes{continuing access to central bank money,} 
  \item improving resilience, 
  \item increasing payments diversity, 
  \item encouraging financial inclusion,
  \item improving cross-border payments,
  \item supporting privacy, and 
  \item facilitating fiscal transfers \cite{bis-gocb-cbdc-designinterop}.
\end{enumerate}

\noindent
The design of a CBDC and its underlying system could potentially lead
to significant risks, ranging from cyber security risks \cite{bis-aer2021-cbdcs}
to financial stability risks \cite{bis-gocb-cbdc-finstab}. \linebreak
In addition, we identify a further risk of fragmentation
in payments markets and retail deposits unless there is interoperability
between CBDCs and existing forms of money.

In the UK, the Bank of England and HM Treasury have established the CBDC
Taskforce to coordinate the exploration of a potential UK CBDC
\changes{as well as}
two external engagement groups \changes{\textendash} the CBDC Engagement Forum and the CBDC
Technology Forum \changes{\textendash} to gather input on \linebreak 
\changes{non-technological and technological}
aspects respectively, of CBDC \cite{boe-cbdc-mainpage}.
\nochange{The Bank of England and HM Treasury} \changes{were also to}
launch a consultation in 2022,
\changes{which set} out their assessment of the case for a UK CBDC \cite{boe-cbdc-consul}.

\changes{In this paper,} we focus on a potential UK CBDC and describe an
illustrative industry architecture based on the Bank of England's
``platform model'' \cite{boe-cbdc-disc-paper}.
Our contribution is the concept of ecosystems that provide a common
programmability layer across both CBDC and commercial bank money
and thereby place both forms of money on a similar footing.
We hope the architecture presented in this paper will stimulate
discussion and look forward to ongoing industry engagement on CBDC.



\section{\changes{Central Bank Digital Currency} Models and Architectures}
\label{sec:cbdc-arch}

Central banks have described, proposed and piloted several models and architectures
for CBDCs.
The Bank for International Settlements (BIS) has described a range of
CBDC architectures including a single-tier ``direct'' architecture,
two-tier ``hybrid'' and ``intermediated'' architectures, and an ``indirect''
architecture \cite{bis-cbdc-archs}.
\changes{The} BIS has also described models for multi-CBDC arrangements to make
cross-border payments more efficient, namely ``compatible'' CBDC systems,
``interlinked'' CBDC systems, and a ``single'' CBDC system \cite{bis-mcbdc-archs}.
The People's Bank of China has initiated a CBDC pilot that uses a two-tier
architecture\changes{,}
with the central bank issuing digital fiat currency to authorised operators
who take charge of exchange and circulation \cite{pboc-ecny}.
The Estonian Central Bank is experimenting with a bill-based CBDC money scheme
built on a partitioned blockchain architecture \cite{ecb-bill-digeuro}.
The Federal Reserve Bank of Boston and the Massachusetts Institute
of Technology have prototyped two CBDC systems with a central transaction
processor, one with an ``atomizer'' architecture and another with a
``two-phase commit'' architecture \cite{fedbos-mit-paper}.

\begin{figure}[!h]
  \captionsetup{width=14cm}
  \begin{center}
  \includegraphics[width=0.6\linewidth]{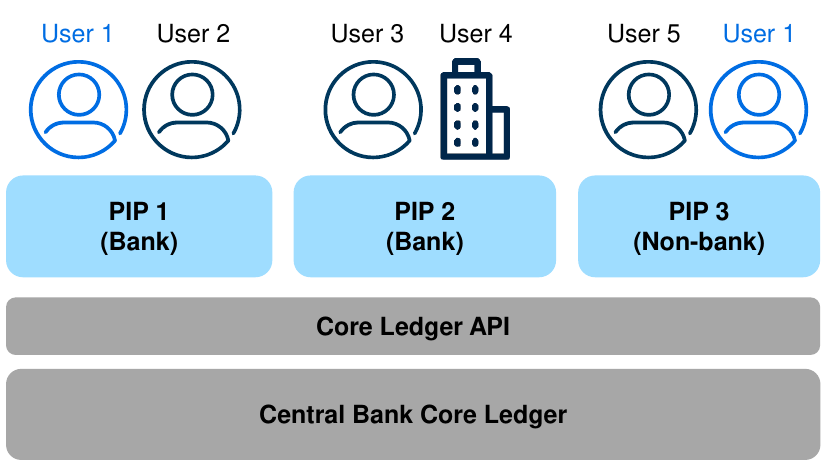}
  \end{center}
  \vspace{-4mm}
  \caption{\footnotesize{\changes{The} Bank of England's ``platform model'' for CBDC
    provision, comprising
    a core ledger, \changes{an} \nochange{application programming interface} (API), 
    \nochange{Payment Interface Providers} (PIPs), and users.
    \nochange{Figure adapted with permission from \cite{boe-cbdc-disc-paper}.}}}
  \label{fig:uk-cbdc-platmodel-fig}

\end{figure}

The Bank of England has described several potential models for CBDC provision
including a ``platform model'', a ``pooled account model'', an
``intermediated token model'', and a ``bearer instrument model'' \cite{boe-models-nov}.
The ``platform model'' (\changes{Figure \ref{fig:uk-cbdc-platmodel-fig}}), which
we adopt in this paper, comprises the Bank of England operating a core
ledger and providing access via application programming interfaces (APIs) 
to authorised and regulated Payment Interface Providers (PIPs) that provide
users with access to CBDC.

\section{Illustrative Industry Architecture}
\label{sec:uk-cbdc-arch}

In this section, we describe an illustrative industry architecture
for a potential UK CBDC by identifying \changes{the} initial requirements, describing the
logical architecture, and analysing how it meets the requirements.
We adopt the Bank of England’s platform model\changes{; that is,} we do not
consider other models in the remainder of this paper.


\subsection{Initial Requirements}
\label{sec:uk-cbdc-arch-kps}

We first identify the following initial requirements as a basis for
developing the architecture:

\begin{itemize}

  \item \emph{Characteristics}: \changes{the architecture should} support the
    characteristics of a CBDC system
    identified by the Bank of England\changes{, namely it should be \cite{boe-designchars-sep}}:
    \begin{enumerate}[(i)]
      \item \changes{reliable and resilient,}
      \item fast and efficient, and
      \item innovative and open to competition. 
    \end{enumerate}

  \item \emph{Technology choices}: \changes{the architecture should align with}
    the Bank of England’s
    current views on potential technology choices on topics such as
    ledger design, privacy, simplicity and programmability  \cite{boe-designchars-sep}.

  \item \emph{Interoperability}: \changes{the architecture should} avoid
    fragmentation by ensuring
    CBDCs and commercial bank money are interoperable and have
    similar operational capabilities.
    
\end{itemize}

\subsection{Logical Architecture} 

\begin{figure}[!h]
  \begin{center}
  \includegraphics[width=0.95\linewidth]{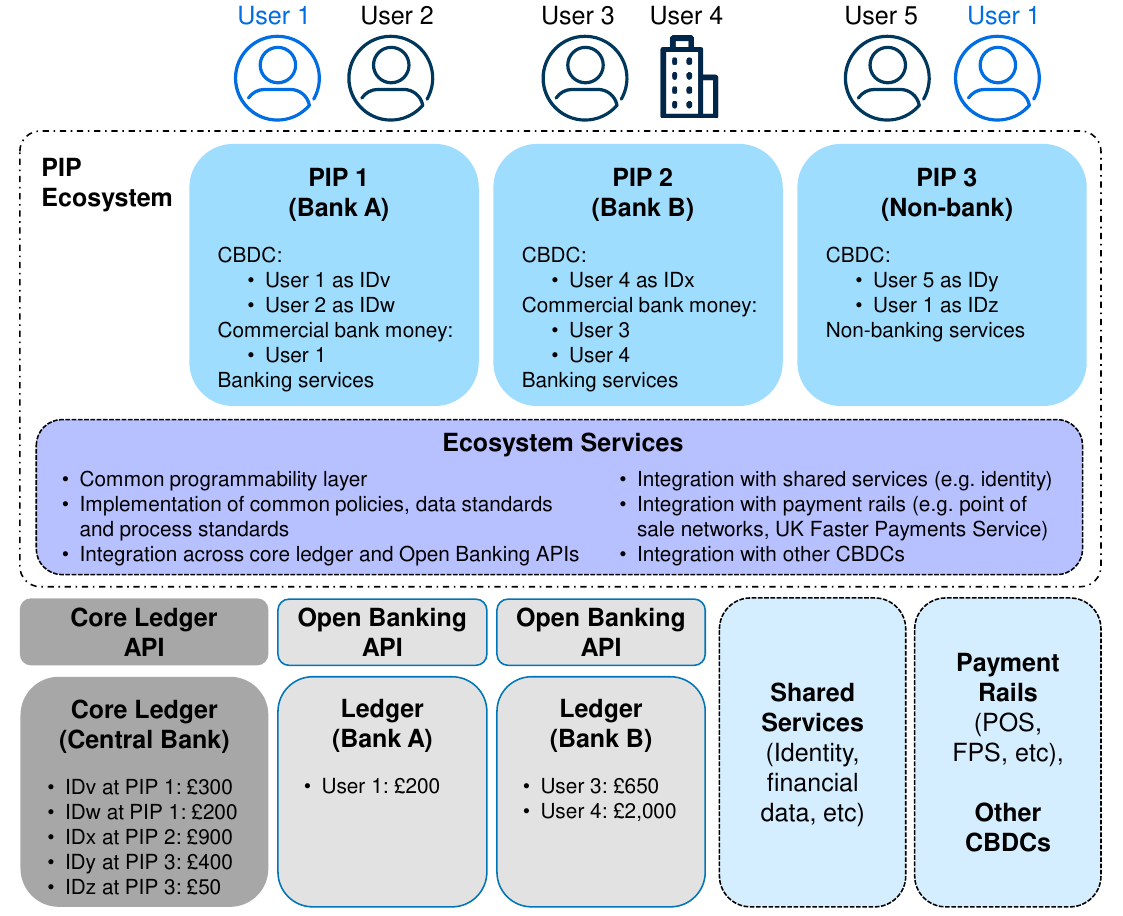}
  \end{center}
  \vspace{-4mm}
  \caption{\footnotesize{An illustrative industry architecture including
  common overlay services across UK CBDC and commercial bank money.
  There could potentially be multiple ecosystems providing competing
    services using different platforms and technologies, but using
    common policies and standards.
  Ecosystem services may be operated by financial market infrastructures.
  \changes{User 1} has accounts at Bank A and the central bank,
  \changes{User 2 has an account} at the central bank,
  \changes{User 3 has an account} at Bank B,
  \changes{User 4 has accounts} at Bank B and the central bank, and
  \changes{User 5 has an account} at the central bank.}}
  
  \label{fig:uk-cbdc-iarch-fig}

\end{figure}

\changes{Figure \ref{fig:uk-cbdc-iarch-fig} depicts the logical architecture.}
\changes{The} key aspects of \changes{this} architecture, which extends the platform model, are
summarised below:

\begin{itemize}

  \item The Bank of England
    \changes{operates the core ledger that records the CBDC
    value and processes the payment transactions made using the CBDC,
    and it provides access to the core ledger via APIs.}
    Users are linked to their CBDC balances and payment transactions
    on the core ledger with pseudonymous identities.
    We highlight that the core ledger APIs could potentially be similar
    to \nochange{Open Banking} APIs \cite{openbanking-apis} with some enhancements
    such as new APIs for opening and closing CBDC accounts.

  \item We introduce PIP ecosystems that provide competing
    services including:
    \begin{enumerate}[(i)]
      \item \changes{the implementation of common policies, data standards and process
        standards,}
      \item integration across the core ledger APIs provided by the Bank
        of England and \nochange{Open Banking} APIs provided by commercial banks,
      \item integration with shared services such as identity providers
        and financial data vendors,
      \item integration with other payment rails such as the UK Faster Payments
        Service (FPS) and point of sale (POS) networks,
      \item integration with other CBDC systems, and
      \item a programmability layer that operates across all of these
        services and provides a foundation for creating new automated
        behaviours and innovative products.
    \end{enumerate}

  \item PIPs are authorised and regulated firms, which can include
    both commercial banks and non-banks.
    They onboard retail and business users, provide customer services,
    and fulfil regulatory requirements such as know-your-customer (KYC)
    and anti-money laundering (AML).
    PIPs also pseudonymise user identity and intermediate user access to the CBDC
    system.
    Note that PIPs could potentially deliver these capabilities by
    leveraging ecosystem services.

\end{itemize}

\subsection{Analysis} 

We now analyse the architecture against the initial requirements and
\changes{summarise our} findings in Table \ref{table:key-principle-analysis}.


\begin{center}
  \begingroup
  \renewcommand{\arraystretch}{1.5}     
  \captionsetup{width=15cm}
    
    \begin{longtable}{ >{\raggedright}p{0.27\textwidth} >{\arraybackslash}p{0.68\textwidth} }
     
     \toprule
     \textbf{Requirements} & \textbf{Analysis} \\  [0.5ex] 
     \midrule
     \endfirsthead

     \multicolumn{2}{r}{\textit{Continued from previous page}} \\
     \toprule
     \textbf{Requirements} & \textbf{Analysis} \\  [0.5ex] 
     \midrule
     \endhead

     \bottomrule
     \multicolumn{2}{r}{\textit{Continued on next page}} \\
     \endfoot

     \bottomrule
     \caption{Summary analysis of architecture
      against initial requirements.}
     \label{table:key-principle-analysis}
     \endlastfoot


      Characteristics: \newline 
        (i) reliable and resilient\changes{,} \newline 
        (ii) fast and efficient\changes{, and} \newline 
        (iii) innovative and open to competition  & 

      (i) The Bank of England can exercise control and
        oversight over the core ledger to ensure it is secure, compliant and private.
        Well-designed platforms can deliver resiliency, scalability and
        availability at the core ledger and ecosystem layers.  \newline
      (ii) The architecture introduces ecosystems as a layer
        between PIPs and the core ledger API, but the overhead of this
        indirection should be minimal and ecosystems can provide
        significant benefits including operational efficiencies. \newline
      (iii) PIP ecosystems can build
        competing and innovative services while using common policies
        and standards to ensure interoperability. \\ 

      Technology choices: \newline
       (i) ledger design\changes{,} \newline
       (ii) privacy\changes{,} \newline
       (iii) simplicity\changes{, and} \newline
       (iv) programmability   & 

      (i) The API-based layered architecture allows the technology choices
        for the core ledger to be generally independent of the
        technology choices for other layers such as the ecosystems. \newline
      (ii) The use of pseudonymous identities ensures only appropriate
        parties are aware of user identities, which facilitates privacy
        while retaining the ability to conduct KYC and AML.  \newline
      (iii) The core ledger infrastructure can be kept relatively simple
        because more complex functionality is provided by PIP ecosystems
        instead. \newline
      (iv) Implementing programmability in the PIP ecosystems layer,
        instead of in the core ledger, should reduce security risks
        and complexity at the core ledger.
        Programs running in PIP ecosystems would leverage the core ledger
        acting as the authoritative data store.
        Each ecosystem could implement programmability using its
        platform of choice.  \\ 

      Interoperability    & 
      The PIP ecosystems would provide common policies, data standards and
      process standards across both CBDC and commercial bank money.
      This would avoid fragmentation by ensuring CBDCs and commercial
      bank money are interoperable and have similar operational
      capabilities. \\

    \end{longtable}

  \endgroup
      
\end{center}


\vspace{-15mm}

\section{Summary and Further Work}
\label{subsec:summary-and-further-work}

This paper \changes{focused} on a potential UK CBDC and presented an
illustrative industry architecture that:

\begin{itemize}

\item adopts and extends the Bank of England's platform model for
  CBDC provision,

\item aligns \changes{with} the Bank of England's currently identified system
  characteristics and views on potential technology choices for
  CBDC infrastructure, and

\item mitigates the risk of fragmentation in payments markets and
  retail deposits by introducing the concept of
  ecosystems that provide a common programmability layer
  across \changes{CBDCs} and commercial bank money.

\end{itemize}

\noindent
Barclays is \changes{developing} a prototype based on the illustrative
industry architecture.
Potential further work includes \changes{the} analysis of any changes needed to 
\nochange{Open Banking} APIs in order to integrate with 
the common programmability layer,
prototyping the specification of the \changes{core ledger} APIs, and elaborating
key customer journeys using the architecture.
We hope this architecture paper will stimulate discussion,
particularly regarding methods to reduce fragmentation risk by placing
CBDCs and commercial bank money on a similar footing, and \changes{we} look
forward to ongoing industry engagement on \changes{CBDCs}.

\vspace{5mm}

\noindent \textbf{Acknowledgements:} 
\noindent
We \changes{thank} Vikram Bakshi (Barclays) for his helpful feedback.

\pretolerance=-1
\tolerance=-1
\emergencystretch=0pt

\bibliography{uk-cbdc-arch-paper}
\bibliographystyle{plain}

\end{document}